\documentclass[aps,prb,twocolumn]{revtex4-2}

\usepackage{enumerate}
\usepackage{amsmath,amsthm, amscd, amssymb, amsfonts}
\usepackage{graphicx}
\setcitestyle{super}
\begin{document}


\title{Noninertial Reference Frames in Special Relativity}



\author{Justo Pastor Lambare}
\email[]{jupalam@gmail.com}

\affiliation{Universidad Nacional de Asunci\'{o}n, FACEN, Ruta Mcal. J. F. Estigarribia, Km 11 Campus de la UNA, San Lorenzo, Paraguay}



\date{\today}

\begin{abstract}
Although the special theory of relativity was conceived to relate inertial reference frames, it is not generally known by beginning students, or even by academics not specializing in general relativity, that it can be applied to noninertial reference frames.
The general perception among non-specialists seems to be that noninertial reference frames can only be described through general relativity.
This situation has led to some controversy regarding the necessity of general relativity to address problems involving accelerated systems, such as the twin paradox and the description of rotating reference frames.
We discuss the controversy and propose that, to avoid the present confusion, the local validity of special relativity in the absence of gravitation, despite the use of noninertial frames, should be explicitly and clearly stated at an early stage of the learning process.
\end{abstract}


\maketitle
%
\section{Introduction}\label{sec:Intro}
In a revealing study, \cite{pPep23} Pepino and Mabile found from a survey of physics faculty and graduate students from more than 22 physics departments of prestigious institutions in the United States and United Kingdom, that only $37\%$ of faculty members and $10\%$ of the graduate students correctly answered \textit{yes} to the question ``Is SR [Special Relativity] capable of describing physics in accelerated reference frames? (\textit{yes or no})''.

The question is not whether special relativity can describe accelerated objects in an inertial reference frame, but whether special relativity can describe spacetime for observers fixed to an accelerated reference frame.
Two classic examples of noninertial reference frames are those fixed to a rotating platform and a uniformly accelerated observer.
The question involves how the spacetime coordinates $(y^0,y^1,y^2,y^3)$ of an inertial observer are related to the spacetime coordinates $(x^0,x^1,x^2,x^3)$ of observers fixed to a noninertial reference frame



In section \ref{sec:TC}, we analyze why there is confussion on this issue.
Section \ref{sec:NIF_G_SR} explains the heuristic approach to interpreting general spacetime coordinates.
Section \ref{sec:FD} presents the formal method that permits the precise implementation of the interpretation explained in section \ref{sec:NIF_G_SR} based on the local validity of special relativity.
Section \ref{sec:ACE} discusses an application clarifying some common misunderstandings existing outside specialist circles.
%
\section{The Controversy}\label{sec:TC}
As often happens when a subject is controversial, the reason lies in the fact that there exist two possible admissible answers depending on the starting premises.

There are two ways to approach general relativity.
One is axiomatically through a rigorous mathematical formulation based on concepts of differential geometry. 
The other is constructively following a path through basic and intuitive physical principles.

The first case employs a top-down approach, starting from axioms, whith special relativity contained as a special case.
In the second, bottom-up approach, general relativity is ``constructed'' from special relativity through heuristic concepts such as the principle of equivalence.

It is the second bottom-up perspective that requires applying special relativity to noninertial frames of reference as an initial step, independently of the existence of a gravitational field, ultimately justifying the name ``general relativity.''
This approach enabled Einstein to predict effects such as the gravitational red shift already in 1907 before developing a comprehensive theory of gravitation.\cite{bPai82}
%
\section{Relationship between special relativity, noninertial reference frames, and gravitation}\label{sec:NIF_G_SR}
Einstein introduced the special theory of relativity in his celebrated 1905 ``June'' paper.\cite{ibEin07}
Although special relativity was initially conceived to deal exclusively with spacetime in inertial coordinates through the Lorentz transformation, in this article, Einstein also included the key idea for the special relativistic treatment of noninertial frames of reference by predicting the behavior of an accelerated clock.

There is no problem with special relativity describing the accelerated motion of the clock as an object.
However, the fundamental question here is can we predict how time passes for an observer attached to an accelerated object?

Since special relativity deals only with inertial observers, to calculate the actual time measured by an accelerated clock, Einstein included the following assumption.\cite{ibEin07}
\begin{quote}
  \textit{``It is at once apparent that this result still holds good if the clock moves from A to B in a polygonal line, and also when the points A and B coincide.\\ 
  If we assume that the result proved for a polygonal line is also valid for a continuously curved line ... ''}
\end{quote}
He made the clock traverse each polygonal side with constant velocity, i.e., he partitioned a continuous, accelerated path into a finite number of rectilinear sections, considering the clock to be at constant velocity through each section.
This enabled Einstein to apply the Lorentz transformation to calculate the clock's instantaneous rate under acceleration. 
In other words, Einstein was approaching accelerated motion by treating it as the limit of infinitesimal segments of constant velocity motion.

Note Einstein's explicit acknowledgment of the assumption through his expression ``\textit{If we assume...}''
From then on, however, he rarely mentioned the assumption explicitly.

Mashhoon appropriately called Einstein's assumption the \textit{``Hypothesis of Locality (HL).''}  \cite{pMas90} 

\begin{figure}[ht]
\includegraphics[scale=0.9,bb=-1cm 4cm 13cm 0]{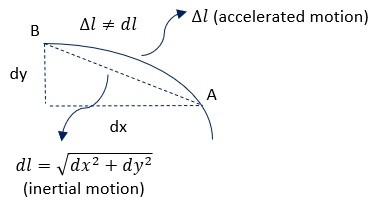}\vspace{3.5cm}
\caption{Differential Pythagorean theorem analogy of Einstein's 1905 assumption. In this analogy, the curved segment represents the actual accelerated motion. The rectilinear segment represents the constant velocity path. While the use of the Pythagorean formula approximates the length of the curved segment, the use of the Lorentz transformation approximates the description of accelerated motion.}
\label{fig1}
\end{figure}

The identification of an accelerated system with a momentarily comoving inertial reference frame\cite{BSch09} allows the local application of the Lorentz transformation. 
It is analogous to the way the approximation of an infinitesimal arc of a curve by a straight segment allows the local application of the Pythagorean theorem; $dl^2=dx^2+dy^2$, although the arc is not a straight line.
In (Fig.\ref{fig1}), $\Delta l$ represents the actual infinitesimal path while $dl$ is the rectilinear approximation.
%
\vspace{.7cm}

\subsection{Einstein Equivalence Principle (EEP) and the HL}
It is the equivalence between acceleration and gravitation that permits the replacement of a gravitational field with an accelerated reference frame. 
This procedure can be interpreted as the application of the equivalence of gravitation and acceleration followed by the HL.

That the local ``infinitesimal'' application of special relativity to accelerated frames is at the heart of general relativity was explained by Einstein himself: \cite{pFletcher23}
\begin{quote}
\textit{The general theory of relativity rests entirely on the premise that each infinitesimal line element of the spacetime manifold physically behaves like the four-dimensional manifold of the special theory of relativity. Thus, there are infinitesimal coordinate systems (inertial systems) with the help of which the $ds$ are to be defined exactly like in the special theory of relativity. The general theory of relativity stands or falls with this interpretation of $ds$. It depends on the latter just as much as Gauss' infinitesimal geometry of surfaces depends on the premise that an infinitesimal surface element behaves metrically like a flat surface element...}
\end{quote}
Although above, Einstein did not mention the motivation for assuming the infinitesimal validity of special relativity,
\footnote{In the modern mathematical formulation of the theory, Einstein's description of the infinitesimal validity of special relativity is expressed by postulating that the tangent space is a Minkowski vector space}
Misner, Thorne, and Wheeler \cite{bMTW73} is more explicit regarding the use of local accelerated reference frames,
\begin{quote}
\textit{A tourist in a powered interplanetary rocket feels ``gravity.'' Can a physicist by local effects convince him that his ``gravity'' is bogus? Never, says Einstein's principle of the local equivalence of gravity and accelerations. But then the physicist will make no errors if he deludes himself into treating true gravity as a local illusion caused by acceleration. Under this delusion, he barges ahead and solves gravitational problems by using special relativity.}
\end{quote}
The key point above is \textit{``But then the physicist will make no errors if he deludes himself into treating true gravity as a local illusion caused by acceleration;''} i.e., he replaces gravitation with acceleration and then applies special relativity ``locally'' in the momentarily inertial reference frame.


We believe this is a deep physical insight that remains hidden in the mathematical formulation. 
Still, the fact that acceleration comes first and gravitation is explained as an outgrowth of acceleration through the equivalence principle is exquisitely explained in the above fable by Misner, Thorne, and Wheeler.
%
\section{Formal Description of the Local Equivalence Between Inertial Motion and Acceleration}\label{sec:FD}
Here we will show explicitly why it is misleading to claim that noninertial reference frames require relativistic gravitation.

The characteristic properties distinguishing the inertial reference frames are the homogeneity and isotropy of space and the homogeneity of time.\cite{bLaL76}
Further, it can be proved that as a consequence of the physical equivalence of all inertial reference frames, their spacetime coordinates transform linearly,\cite{pBer_Gor69}, and the differential interval, 
\begin{eqnarray}\label{eq:Mm}
ds^2&=&(y^0)^2-(y^1)^2-(y^1)^2-(y^1)^3\\
    &=&\sum_{\alpha,\beta=0}^3\eta_{\alpha\beta}dy^\alpha dy^\beta
\end{eqnarray}
where $y^0=ct$, remains invariant under the Lorentz transformation.\cite{bLaL75}
The homogeneity and isotropy properties are also reflected in the fact that $\eta_{\alpha\beta}=diag(+1,-1,-1,-1)$ remains constant throughout space and time, where
\begin{equation}
diag(+1,-1,-1,-1)=
\left(
\begin{array}{cccc}
1 & 0 & 0 & 0\\
0 & -1 & 0 & 0\\
0 &  0 &-1 & 0\\
0 &  0 & 0 &-1\\
\end{array}
\right)
\end{equation}
When we admit noninertial reference frames, the homogeneity and isotropy that single out the inertial systems are lost, despite remaining in flat Minkowski spacetime.
As a consequence, the coordinate transformations are no longer linear.

Since special relativity is still locally valid, the differential interval $ds$ remains invariant.

When (\ref{eq:Mm}) is expressed in non-inertial coodinates by
\begin{eqnarray}
   y^\alpha &=& y^\alpha(x^0,x^1,x^2,x^3)\\
  dy^\alpha &=& \sum_{\mu=0}^3\frac{\partial y^\alpha}{\partial x^\mu}dx^\mu\\
  ds^2      &=& \sum_{\mu,\nu=0}^3\,\underbrace{\eta_{\alpha\beta}\frac{\partial y^\alpha}{\partial x^\mu}\frac{\partial y^\beta}{\partial x^\nu}}_{g_{\mu\nu}}\,dx^\mu dx^\nu\\
  ds^2      &=& \sum_{\mu,\nu=0}^3g_{\mu\nu}\,dx^\mu dx^\nu\label{eq:gm}
\end{eqnarray}
where $x^0=ct$ and $g_{\mu\nu}(x^0,x^1,x^2,x^3)$ is not necessarily iqual to $\eta_{\mu\nu}$.
\footnote{In this case, the fact that $g_{\mu\nu}\neq\eta_{\mu\nu}$ does not mean spacetime becomes curved. It means that we are using a noninertial reference frame to describe flat spacetime.}

The price we have to pay for this generality is that the spacetime coordinates $x^\mu$ have a significance only locally through the physical interpretation of the interval $ds$.
To extract this meaning from (\ref{eq:gm}) at a given spacetime point, we must put it in the same form as (\ref{eq:Mm}) at that point, as allowed by the HL.


Separating the spatial part in the expression of the interval we have \cite{pLog88}
{\small
\begin{eqnarray}
  ds^2 &=&\sum_{\mu,\nu=0}^3 g_{\mu\nu}dx^\mu dx^\nu\nonumber\\
       &=& g_{00}(dx^0)^2 + \sum_{k=1}^3 2g_{0k}dx^0 dx^k + \sum_{i,k=1}^3 g_{ik}dx^i dx^k\label{eq:gm1}\nonumber\\
       &=& g_{00}\left((dx^0)^2 + \sum_{k=1}^3 2\frac{g_{0k}}{g_{00}}dx^0 dx^k\right) + \sum_{i,k=1}^3 g_{ik}dx^i dx^k\nonumber\\
       &=& g_{00}\left((dx^0)^2 + \sum_{k=1}^3 2\frac{g_{0k}}{g_{00}}dx^0 dx^k +\sum_{i,k=1}^3\frac{g_{0i}g_{0k}}{g_{00}^2}dx^i dx^k\right.\nonumber\\
       & & \left.-\sum_{i,k=1}^3\frac{g_{0i}g_{0k}}{g_{00}^2}dx^i dx^k\right) + \sum_{i,k=1}^3 g_{ik}dx^i dx^k \label{eq:pmterm}
\end{eqnarray}
}
%
which can be rearranged as
\begin{widetext}
\begin{eqnarray}
   ds^2  &=&g_{00}\left(dx^0 + \sum_{k=1}^3\tfrac{g_{0k}}{g_{00}}dx^k\right)^2 -\sum_{ik=1}^3 \underbrace{\left(-g_{ik}+\tfrac{g_{0i} g_{0k}}{g_{00}}\right)}_{\gamma_{ik}}dx^i dx^k  \label{eq:gm2}
\end{eqnarray}
\end{widetext}
Expressing the last equation in ordinary local inertial coordinates we have
%
\begin{eqnarray}
 ds^2&=&(y^0)^2-(y^1)^2-(y^1)^2-(y^1)^3\\
     &=& c^2d\tau^2-dl^2\label{eq:ni} 
\end{eqnarray}
%
Comparing (\ref{eq:gm2}) with (\ref{eq:ni}), we find that the spatial distance is given by,
\begin{eqnarray}
  dl^2 &=& \gamma_{ik}dx^idx^k\label{eq:edd} 
\end{eqnarray}
and the actual elapsed time, or proper time, measured by physical clocks is,
%
\begin{eqnarray}
  d\tau^2 &=& \frac{g_{00}}{c^2}\left(dx^0 +\sum_{k=1}^3\tfrac{g_{0k}}{g_{00}}dx^k\right)^2\label{eq:pts}
\end{eqnarray}

It is also possible to obtain (\ref{eq:edd}) and (\ref{eq:pts}) by exchanging light signals between the spatial positions $x^k$ and $x^k+dx^k$ (cf. Ref. \onlinecite{bLaL75}). 

When we want the elapsed time at a point at a fixed location in the $x^k$ coordinates, we must put $dx^k=0$  in (\ref{eq:pts}),
%
\begin{eqnarray}\label{eq:dt}
  d\tau &=& \frac{\sqrt{g_{00}}}{c}dx^0
\end{eqnarray}
%
For a finite time interval marked by a clock at a fixed value of $x^i$, we have,
\begin{eqnarray}\label{eq:dti}
  \tau_2-\tau_1 &=& \int_{x^0_{(1)}}^{x^0_{(2)}}\frac{\sqrt{g_{00}}}{c}dx^0
\end{eqnarray}

Note that the only hypothesis necessary to extract the actual space and time information contained in the metric tensor $g_{\mu\nu}$ is the HL, which justifies equating (\ref{eq:gm2}) and (\ref{eq:ni}); gravitation plays no role in the procedure.
The full structure of general relativity is necessary only to find how the mass-energy distribution determines the $g_{\mu\nu}$, not its local space and time interpretation.
%
%
\section{A Clarifying Example}\label{sec:ACE}
The incorrect relativistic analysis of rotating noninertial reference frames has led to some erroneous disproofs of special relativity.
\footnote{In this section, we are assuming a basic familiarity with the Sagnac effect. Clarifying discussions of the peculiarities of spacetime in rotating reference frames can also be found in Refs. \cite{bMol52,pWeb97,pWeb99}}

One such purported disproof is known as Selleri's paradox \cite{pSel97,pKas12}, and another related one is the incorrect relativistic analysis of the Sagnac effect.\cite{pEng15,pKip21a,pSpa21}
In both cases, the mistake consists of erroneously assuming global properties of homogeneity and isotropy that are only valid in inertial reference frames and the dismissal of the ``only local'' validity of special relativity.

A concrete flawed argument against special relativity based on the Sagnac effect claims the following:\cite{pEng15,pKip21a}
\begin{enumerate}[a)]
  \item The special relativistic addition of velocities predicts that light speed is invariant for observers fixed along the rim of the rotating platform. (Correct.)
  \item Since the clockwise and counterclockwise light beams traverse symmetric paths at the same speeds, an experimenter at $C$ (Fig. \ref{fig2}) rotating with the interferometer must observe both beams arriving simultaneously. Hence, special relativity predicts no displacement of the interference fringes, thus contradicting empirical observation.  (Incorrect.)
\end{enumerate}

%
\subsection{The Special Relativistic Solution}\label{subsec:TSRS}
For the case that concerns us in this section, i.e., the prediction of the Sagnac effect from within the rotating platform, we transform from the coordinates in the inertial reference frame with origin in the disc center $(ct',r',\phi',z')$ to general noninertial coordinates fixed in the rotating platform $(ct,r,\phi,z)$,
\begin{equation}\label{eq:trans}
t=t',\qquad r=r', \qquad\phi=\phi'-\Omega t, \qquad z=z'
\end{equation}
Here we remark on a usual misinterpretation.
The time transformation $t=t'$ is sometimes misinterpreted as a Galilean transformation of time, meaning that the variable $t$ inside the platform represents the real time as measured by physical clocks at rest with respect to the rotating platform. \cite{pEng15,pKip21a}

However, that is an incorrect interpretation of the time coordinate.
When expressed in terms of an arbitrary (noninertial) reference frame, the time coordinate is a mere parameter used to calculate the actual proper time measured by physical clocks in that frame.

Although according to (\ref{eq:trans}), the time coordinate $t$ can be interpreted as the time of the background inertial reference frame, inside the platform it is only an auxiliary parameter that allows the calculation of the true time indicated by physical clocks attached to the rotating platform according to (\ref{eq:dt}) and (\ref{eq:dti}).
This interpretation is not some esoteric effect coming from general relativity but the mere consequence of a consistent ``local'' application of special relativity. 

For the differential interval $ds$, (\ref{eq:trans}) gives,
\begin{equation}\label{eq:ri} 
ds^2=\underbrace{\left(1-\beta^2\right)}_{g_{00}} c^2dt^2-dr^2-r^2d\phi^2-dz^2-2\beta rc\,dt\,d\phi
\end{equation}
where $\beta=v_0/c$, $v_0=r\Omega$, and $x^0=ct,\,x^1=r,\,x^2=\phi,\,x^3=z$.
The time elapsed on a physical clock at a fixed position on the rotating platform is obtained from (\ref{eq:dti}) and (\ref{eq:ri}),
\begin{eqnarray}
\tau_{2}-\tau_{1} &=&\int_{x^0_{1}}^{x^0_{2}}\frac{\sqrt{g_{00}}}{c}\,dx^0\label{eq:fgtau}\\ 
\tau_{2}-\tau_{1} &=&\int_{t_{1}}^{t_{2}}\frac{dt}{\gamma}\label{eq:fctau} 
\end{eqnarray}
where $\gamma=1/\sqrt{1-\beta^2}$.
To evaluate the Sagnac effect, we need to calculate the elapsed times $\tau^+$ and $\tau^-$ in the location of the interferometer.

According to the HL, the light paths must satisfy $ds=0$, which for $r=R$ (Fig.\ref{fig2}) leads to,
{\small
\begin{eqnarray}
dt^+     &=& +\frac{R d\phi^+}{c(1-\beta)}\,,\qquad d\phi^+>0,\,\text{(forward beam)}\label{eq:ngp}\\ 
dt^-     &=& -\frac{R d\phi^-}{c(1+\beta)}\,,\qquad d\phi^-<0,\,\text{(backward beam)}\label{eq:ngn}   
\end{eqnarray}
}
Replacing (\ref{eq:ngp}) in (\ref{eq:fctau}),
\begin{eqnarray}
\tau^+ &=&\int_0^{2\pi}\frac{1}{\gamma}\frac{R d\phi^+}{c(1-\beta)}\\
       &=&\frac{1}{\gamma}\frac{L}{c(1-\beta)}
\end{eqnarray}
Analogously, from (\ref{eq:ngn}) and (\ref{eq:fctau}),
\begin{eqnarray}
\tau^- &=&\int_0^{-2\pi}-\frac{1}{\gamma}\frac{R d\phi^-}{c(1+\beta)}\\
       &=&\frac{1}{\gamma}\frac{L}{c(1+\beta)}
\end{eqnarray}
The displacement of the interference fringes is determined by the difference,
\begin{eqnarray}
\tau^+-\tau^- &=&\frac{2Lv_0}{c^2\sqrt{1-(\frac{v_0}{c})^2}} 
\end{eqnarray}
Putting $L=2\pi R $ and $v_0=\Omega R$, the last result, to first order in the velocity, gives the classical value of the Sagnac effect
\begin{eqnarray}
\tau^+-\tau^- &=&\frac{4A\Omega}{c^2}
\end{eqnarray}
where $A=\pi R^2$.

That special relativity suffices to explain the Sagnac effect from within the rotating platform is also highlighted
in Ref. \cite{pLog88}
%
\subsection{Further Clarifications}\label{subsec:FC}
Although in Section \ref{subsec:TSRS}, we showed how to extract the information contained in the metric solving the problem of the Sagnac effect according to a consistent local application of special relativity, point b) at the beginning of Section \ref{sec:ACE} may still seem puzzling.

How is it possible that after following symmetrical paths traversed at the same speeds, the two beams do not arrive simultaneously?
As we previously observed, in noninertial reference frames, the global homogeneity and isotropy of space, as well as the homogeneity of time, are no longer valid.
We can assert that the local-only validity of special relativity in such cases results in even more mind-boggling counterintuitive repercussions than when its global validity applies.

\begin{figure}[t]
\includegraphics[scale=0.4,bb=4.5cm 12cm 15cm 0]{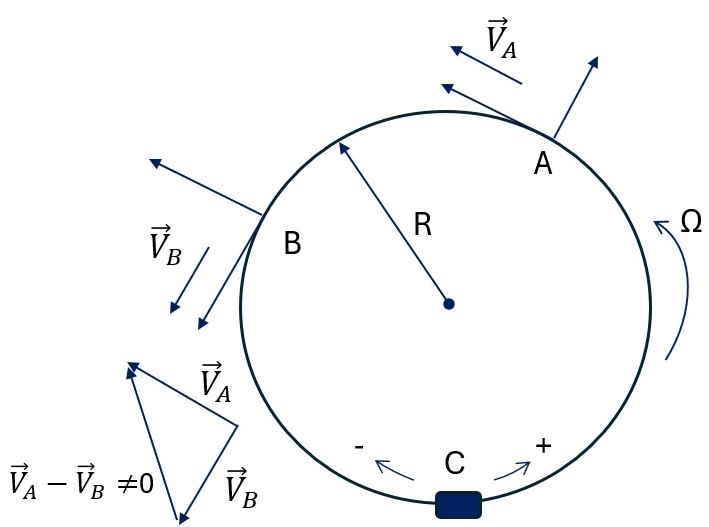}\vspace{5cm}
\caption{Momentarily comoving inertial reference frames at points A and B.}
\label{fig2}
\end{figure}

The fact that two observers, such as $A$ and $B$, fixed to the platform (Fig. \ref{fig2}), are instantaneously at rest in different inertial frames and are not themselves at rest with respect to each other has unexpected consequences.
Also, in the concrete case of a rotating platform, the rotation has a determinate direction that breaks the symmetry of the two paths.
\vspace{4cm}
Furthermore, the mean speed of the beams calculated with one fixed clock at C does not coincide with the constant speed $c$ measured with local clocks placed along the beam paths.
A similar effect also takes place in the gravitational field of the sun.
It is known as the Shapiro effect and is considered the fourth classical test of general relativity. \cite{pSha64}
This analogy is also mentioned in Ref. \onlinecite{pBen19}

Additionally, the subtlety that the equivalence between acceleration and inertial motion is only local renders the EEP a local principle as well, which cannot be applied globally.
Dismissing this fact can easily lead to incorrect inferences about the consistency and validity of the theory of relativity.\cite{pVav24,pLam25b}

An instructive and interesting instance discussing similar issues can be found in an exchange between Robert D. Klauber and T. A. Weber. \cite{pKla99,pWeb99}
%
\section{Conclusions}
Einstein recognized in 1905 that the equivalence between an accelerated reference frame and a momentarily comoving inertial system constitutes an assumption.
However, in later applications of the method, Einstein omitted any explicit reference to the assumption and applied it systematically without further mention or clarification.

Except for the particular case of the HC known as the \textit{clock hypothesis}, which some physicists and philosophers admit as an independent hypothesis,
\footnote{The fact that some find the \textit{clock hypothesis} an independent assumption, \cite{bFoc66,pRos78,bTor96} while others a consequence of the theory that transforms it into a theorem \cite{pSfarti17,pFletcher13} is an example of the duality between considering the HL an independent assumption or a natural consequence of the theory.}
the local application of special relativity to accelerated systems of reference has become a self-evident procedure that does not require further clarification.
The principle issues underlying confusion are:
\begin{itemize}
  \item An accelerated system is not an inertial one even though its coordinate axes may instantaneously coincide with an inertial system.
  \item The special theory of relativity was originally formulated for relating space and time only between inertial frames of reference.
\end{itemize}
The perplexity among nonspecialists arising from applying special relativity to accelerated reference frames generally arises from these two issues.

Thus, whether accepted as an independent hypothesis or considered as an evident or even mathematical consequence of the theory, to avoid the confusion revealed by Pepino and Mabile, students should be explicitly warned about the local validity of special relativity in noninertial frames of reference at an early stage of exposure to special relativity.

\section*{Author Declarations}
\subsection*{Conflict of Interest}
The author has no conflict of interest to disclose.

\bibliography{zRelativity}

\begin{thebibliography}{33}%
\makeatletter
\providecommand \@ifxundefined [1]{%
 \@ifx{#1\undefined}
}%
\providecommand \@ifnum [1]{%
 \ifnum #1\expandafter \@firstoftwo
 \else \expandafter \@secondoftwo
 \fi
}%
\providecommand \@ifx [1]{%
 \ifx #1\expandafter \@firstoftwo
 \else \expandafter \@secondoftwo
 \fi
}%
\providecommand \natexlab [1]{#1}%
\providecommand \enquote  [1]{``#1''}%
\providecommand \bibnamefont  [1]{#1}%
\providecommand \bibfnamefont [1]{#1}%
\providecommand \citenamefont [1]{#1}%
\providecommand \href@noop [0]{\@secondoftwo}%
\providecommand \href [0]{\begingroup \@sanitize@url \@href}%
\providecommand \@href[1]{\@@startlink{#1}\@@href}%
\providecommand \@@href[1]{\endgroup#1\@@endlink}%
\providecommand \@sanitize@url [0]{\catcode `\\12\catcode `\$12\catcode
  `\&12\catcode `\#12\catcode `\^12\catcode `\_12\catcode `\%12\relax}%
\providecommand \@@startlink[1]{}%
\providecommand \@@endlink[0]{}%
\providecommand \url  [0]{\begingroup\@sanitize@url \@url }%
\providecommand \@url [1]{\endgroup\@href {#1}{\urlprefix }}%
\providecommand \urlprefix  [0]{URL }%
\providecommand \Eprint [0]{\href }%
\providecommand \doibase [0]{https://doi.org/}%
\providecommand \selectlanguage [0]{\@gobble}%
\providecommand \bibinfo  [0]{\@secondoftwo}%
\providecommand \bibfield  [0]{\@secondoftwo}%
\providecommand \translation [1]{[#1]}%
\providecommand \BibitemOpen [0]{}%
\providecommand \bibitemStop [0]{}%
\providecommand \bibitemNoStop [0]{.\EOS\space}%
\providecommand \EOS [0]{\spacefactor3000\relax}%
\providecommand \BibitemShut  [1]{\csname bibitem#1\endcsname}%
\let\auto@bib@innerbib\@empty
\bibitem [{\citenamefont {Pepino}\ and\ \citenamefont {Mabile}(2023)}]{pPep23}%
  \BibitemOpen
  \bibfield  {author} {\bibinfo {author} {\bibfnamefont {R.~A.}\ \bibnamefont
  {Pepino}}\ and\ \bibinfo {author} {\bibfnamefont {R.~W.}\ \bibnamefont
  {Mabile}},\ }\bibfield  {title} {\bibinfo {title} {{A {M}isconception
  {R}egarding the {E}instein {E}quivalence {P}rinciple and a {P}ossible {C}ure
  {U}sing the {T}win {P}aradox}},\ }\href {https://doi.org/10.1119/5.0075153}
  {\bibfield  {journal} {\bibinfo  {journal} {The Physics Teacher}\ }\textbf
  {\bibinfo {volume} {61}},\ \bibinfo {pages} {118} (\bibinfo {year}
  {2023})}\BibitemShut {NoStop}%
\bibitem [{\citenamefont {Pais}(1982)}]{bPai82}%
  \BibitemOpen
  \bibfield  {author} {\bibinfo {author} {\bibfnamefont {A.}~\bibnamefont
  {Pais}},\ }\href@noop {} {\emph {\bibinfo {title} {Subtle is the {L}ord}}}\
  (\bibinfo  {publisher} {Oxford University Press},\ \bibinfo {address}
  {London},\ \bibinfo {year} {1982})\BibitemShut {NoStop}%
\bibitem [{\citenamefont {Einstein}(2007)}]{ibEin07}%
  \BibitemOpen
  \bibfield  {author} {\bibinfo {author} {\bibfnamefont {A.}~\bibnamefont
  {Einstein}},\ }\bibinfo {title} {On the {E}lectrodynamics of {M}oving
  {B}odies},\ in\ \href@noop {} {\emph {\bibinfo {booktitle} {A {S}tubbornly
  {P}ersistent {I}llusion}}},\ \bibinfo {editor} {edited by\ \bibinfo {editor}
  {\bibfnamefont {S.}~\bibnamefont {Hawking}}}\ (\bibinfo  {publisher} {Running
  Press},\ \bibinfo {year} {2007})\ pp.\ \bibinfo {pages} {4--31}\BibitemShut
  {NoStop}%
\bibitem [{\citenamefont {Mashhoon}(1990)}]{pMas90}%
  \BibitemOpen
  \bibfield  {author} {\bibinfo {author} {\bibfnamefont {B.}~\bibnamefont
  {Mashhoon}},\ }\bibfield  {title} {\bibinfo {title} {The hypothesis of
  locality in relativistic physics},\ }\href
  {https://doi.org/https://doi.org/10.1016/0375-9601(90)90670-J} {\bibfield
  {journal} {\bibinfo  {journal} {Physics Letters A}\ }\textbf {\bibinfo
  {volume} {145}},\ \bibinfo {pages} {147} (\bibinfo {year}
  {1990})}\BibitemShut {NoStop}%
\bibitem [{\citenamefont {Schutz}(2009)}]{BSch09}%
  \BibitemOpen
  \bibfield  {author} {\bibinfo {author} {\bibfnamefont {B.}~\bibnamefont
  {Schutz}},\ }\href
  {http://www.amazon.ca/exec/obidos/redirect?tag=citeulike09-20\&amp;path=ASIN/0521887054
  http://www.amazon.de/exec/obidos/redirect?tag=citeulike01-21\&amp;path=ASIN/0521887054
  http://www.amazon.fr/exec/obidos/redirect?tag=citeulike06-21\&amp;path=ASIN/0521887054
  http://www.amazon.jp/exec/obidos/ASIN/0521887054
  http://www.amazon.co.uk/exec/obidos/ASIN/0521887054/citeulike00-21
  http://www.amazon.com/exec/obidos/redirect?tag=citeulike07-20\&path=ASIN/0521887054
  http://www.worldcat.org/isbn/0521887054 htt} {\emph {\bibinfo {title} {{A
  First Course in General Relativity}}}},\ \bibinfo {edition} {2nd}\ ed.\
  (\bibinfo  {publisher} {Cambridge University Press},\ \bibinfo {year}
  {2009})\BibitemShut {NoStop}%
\bibitem [{\citenamefont {Fletcher}\ and\ \citenamefont
  {Weatherall}(2023)}]{pFletcher23}%
  \BibitemOpen
  \bibfield  {author} {\bibinfo {author} {\bibfnamefont {S.~C.}\ \bibnamefont
  {Fletcher}}\ and\ \bibinfo {author} {\bibfnamefont {J.~O.}\ \bibnamefont
  {Weatherall}},\ }\bibfield  {title} {\bibinfo {title} {The {L}ocal {V}alidity
  of {S}pecial {R}elativity, {P}art 1: {G}eometry},\ }\bibfield  {journal}
  {\bibinfo  {journal} {Philosophy of Physics}\ }\href
  {https://doi.org/10.31389/pop.6} {10.31389/pop.6} (\bibinfo {year}
  {2023})\BibitemShut {NoStop}%
\bibitem [{Note1()}]{Note1}%
  \BibitemOpen
  \bibinfo {note} {In the modern mathematical formulation of the theory,
  Einstein's description of the infinitesimal validity of special relativity is
  expressed by postulating that the tangent space is a Minkowski vector
  space}\BibitemShut {NoStop}%
\bibitem [{\citenamefont {Misner}\ \emph {et~al.}(1973)\citenamefont {Misner},
  \citenamefont {Thorne},\ and\ \citenamefont {Wheeler}}]{bMTW73}%
  \BibitemOpen
  \bibfield  {author} {\bibinfo {author} {\bibfnamefont {C.}~\bibnamefont
  {Misner}}, \bibinfo {author} {\bibfnamefont {K.}~\bibnamefont {Thorne}},\
  and\ \bibinfo {author} {\bibfnamefont {J.}~\bibnamefont {Wheeler}},\
  }\href@noop {} {\emph {\bibinfo {title} {Gravitaion}}}\ (\bibinfo
  {publisher} {W. H. Freeman And Company},\ \bibinfo {address} {New York,
  USA},\ \bibinfo {year} {1973})\BibitemShut {NoStop}%
\bibitem [{\citenamefont {Landau}\ and\ \citenamefont
  {Lifshitz}(1976)}]{bLaL76}%
  \BibitemOpen
  \bibfield  {author} {\bibinfo {author} {\bibfnamefont {L.}~\bibnamefont
  {Landau}}\ and\ \bibinfo {author} {\bibfnamefont {E.}~\bibnamefont
  {Lifshitz}},\ }\href@noop {} {\emph {\bibinfo {title} {Course of
  {T}heoretical {P}hysics, {V}olume 1, {M}echanics}}},\ \bibinfo {edition}
  {3rd}\ ed.\ (\bibinfo  {publisher} {ELSEVIER},\ \bibinfo {address} {Oxford,
  UK},\ \bibinfo {year} {1976})\BibitemShut {NoStop}%
\bibitem [{\citenamefont {Berzi}\ and\ \citenamefont
  {Gorini}(1969)}]{pBer_Gor69}%
  \BibitemOpen
  \bibfield  {author} {\bibinfo {author} {\bibfnamefont {V.}~\bibnamefont
  {Berzi}}\ and\ \bibinfo {author} {\bibfnamefont {V.}~\bibnamefont {Gorini}},\
  }\bibfield  {title} {\bibinfo {title} {{Reciprocity {P}rinciple and the
  {L}orentz {T}ransformations}},\ }\href {https://doi.org/10.1063/1.1665000}
  {\bibfield  {journal} {\bibinfo  {journal} {Journal of Mathematical Physics}\
  }\textbf {\bibinfo {volume} {10}},\ \bibinfo {pages} {1518} (\bibinfo {year}
  {1969})}\BibitemShut {NoStop}%
\bibitem [{\citenamefont {Landau}\ and\ \citenamefont
  {Lifshitz}(1975)}]{bLaL75}%
  \BibitemOpen
  \bibfield  {author} {\bibinfo {author} {\bibfnamefont {L.}~\bibnamefont
  {Landau}}\ and\ \bibinfo {author} {\bibfnamefont {E.}~\bibnamefont
  {Lifshitz}},\ }\href@noop {} {\emph {\bibinfo {title} {Course of
  {T}heoretical {P}hysics, {V}olume 2, {T}he {C}lassical {T}heory of
  {F}ields}}},\ \bibinfo {edition} {4th}\ ed.\ (\bibinfo  {publisher}
  {ELSEVIER},\ \bibinfo {address} {Oxford, UK},\ \bibinfo {year}
  {1975})\BibitemShut {NoStop}%
\bibitem [{Note2()}]{Note2}%
  \BibitemOpen
  \bibinfo {note} {In this case, the fact that $g_{\mu \nu }\protect \neq \eta
  _{\mu \nu }$ does not mean spacetime becomes curved. It means that we are
  using a noninertial reference frame to describe flat spacetime.}\BibitemShut
  {Stop}%
\bibitem [{\citenamefont {Logunov}\ and\ \citenamefont
  {Chugreev}(1988)}]{pLog88}%
  \BibitemOpen
  \bibfield  {author} {\bibinfo {author} {\bibfnamefont {A.~A.}\ \bibnamefont
  {Logunov}}\ and\ \bibinfo {author} {\bibfnamefont {Y.~V.}\ \bibnamefont
  {Chugreev}},\ }\bibfield  {title} {\bibinfo {title} {Special theory of
  relativity and the {S}agnac effect},\ }\href
  {https://doi.org/10.1070/PU1988v031n09ABEH005624} {\bibfield  {journal}
  {\bibinfo  {journal} {Soviet Physics Uspekhi}\ }\textbf {\bibinfo {volume}
  {31}},\ \bibinfo {pages} {861} (\bibinfo {year} {1988})}\BibitemShut
  {NoStop}%
\bibitem [{Note3()}]{Note3}%
  \BibitemOpen
  \bibinfo {note} {In this section, we are assuming a basic familiarity with
  the Sagnac effect. Clarifying discussions of the peculiarities of spacetime
  in rotating reference frames can also be found in Refs. \cite
  {bMol52,pWeb97,pWeb99}}\BibitemShut {NoStop}%
\bibitem [{\citenamefont {Selleri}(1997)}]{pSel97}%
  \BibitemOpen
  \bibfield  {author} {\bibinfo {author} {\bibfnamefont {F.}~\bibnamefont
  {Selleri}},\ }\bibfield  {title} {\bibinfo {title} {Noninvariant one-way
  speed of light and locally equivalent reference frames},\ }\href
  {https://doi.org/https://doi.org/10.1007/BF02764121} {\bibfield  {journal}
  {\bibinfo  {journal} {Foundations of Physics Letters}\ }\textbf {\bibinfo
  {volume} {10}},\ \bibinfo {pages} {73} (\bibinfo {year} {1997})}\BibitemShut
  {NoStop}%
\bibitem [{\citenamefont {Kassner}(2012)}]{pKas12}%
  \BibitemOpen
  \bibfield  {author} {\bibinfo {author} {\bibfnamefont {K.}~\bibnamefont
  {Kassner}},\ }\bibfield  {title} {\bibinfo {title} {{Ways to resolve
  Selleri's paradox}},\ }\href {https://doi.org/10.1119/1.4755950} {\bibfield
  {journal} {\bibinfo  {journal} {American Journal of Physics}\ }\textbf
  {\bibinfo {volume} {80}},\ \bibinfo {pages} {1061} (\bibinfo {year}
  {2012})}\BibitemShut {NoStop}%
\bibitem [{\citenamefont {Engelhardt}(2015)}]{pEng15}%
  \BibitemOpen
  \bibfield  {author} {\bibinfo {author} {\bibfnamefont {W.}~\bibnamefont
  {Engelhardt}},\ }\bibfield  {title} {\bibinfo {title} {Classical and
  {R}elativistic {D}erivation of the {S}agnac {E}ffect},\ }\href
  {https://doi.org/https://doi.org/10.48550/arXiv.1404.4075} {\bibfield
  {journal} {\bibinfo  {journal} {Annales de la Fondation Louis de Broglie}\
  }\textbf {\bibinfo {volume} {40}},\ \bibinfo {pages} {149} (\bibinfo {year}
  {2015})}\BibitemShut {NoStop}%
\bibitem [{\citenamefont {Kipreos}\ and\ \citenamefont
  {Balachandran}(2021)}]{pKip21a}%
  \BibitemOpen
  \bibfield  {author} {\bibinfo {author} {\bibfnamefont {E.~T.}\ \bibnamefont
  {Kipreos}}\ and\ \bibinfo {author} {\bibfnamefont {R.~S.}\ \bibnamefont
  {Balachandran}},\ }\bibfield  {title} {\bibinfo {title} {Assessment of the
  relativistic rotational transformations},\ }\href
  {https://doi.org/10.1142/S0217732321501133} {\bibfield  {journal} {\bibinfo
  {journal} {Modern Physics Letters A}\ }\textbf {\bibinfo {volume} {36}},\
  \bibinfo {pages} {2150113} (\bibinfo {year} {2021})}\BibitemShut {NoStop}%
\bibitem [{\citenamefont {Spavieri}\ \emph {et~al.}(2021)\citenamefont
  {Spavieri}, \citenamefont {Gillies},\ and\ \citenamefont {Haug}}]{pSpa21}%
  \BibitemOpen
  \bibfield  {author} {\bibinfo {author} {\bibfnamefont {G.}~\bibnamefont
  {Spavieri}}, \bibinfo {author} {\bibfnamefont {G.~T.}\ \bibnamefont
  {Gillies}},\ and\ \bibinfo {author} {\bibfnamefont {E.~G.}\ \bibnamefont
  {Haug}},\ }\bibfield  {title} {\bibinfo {title} {The {S}agnac effect and the
  role of simultaneity in relativity theory},\ }\href
  {https://doi.org/10.1080/09500340.2021.1887384} {\bibfield  {journal}
  {\bibinfo  {journal} {Journal of Modern Optics}\ }\textbf {\bibinfo {volume}
  {68}},\ \bibinfo {pages} {202} (\bibinfo {year} {2021})}\BibitemShut
  {NoStop}%
\bibitem [{\citenamefont {Shapiro}(1964)}]{pSha64}%
  \BibitemOpen
  \bibfield  {author} {\bibinfo {author} {\bibfnamefont {I.~I.}\ \bibnamefont
  {Shapiro}},\ }\bibfield  {title} {\bibinfo {title} {Fourth {T}est of
  {G}eneral {R}elativity},\ }\href {https://doi.org/10.1103/PhysRevLett.13.789}
  {\bibfield  {journal} {\bibinfo  {journal} {Phys. Rev. Lett.}\ }\textbf
  {\bibinfo {volume} {13}},\ \bibinfo {pages} {789} (\bibinfo {year}
  {1964})}\BibitemShut {NoStop}%
\bibitem [{\citenamefont {Benedetto}\ \emph {et~al.}(2019)\citenamefont
  {Benedetto}, \citenamefont {Feleppa}, \citenamefont {F.}, \citenamefont
  {Licata}, \citenamefont {I.}, \citenamefont {Moradpour},\ and\ \citenamefont
  {H.}}]{pBen19}%
  \BibitemOpen
  \bibfield  {author} {\bibinfo {author} {\bibfnamefont {E.}~\bibnamefont
  {Benedetto}}, \bibinfo {author} {\bibnamefont {Feleppa}}, \bibinfo {author}
  {\bibnamefont {F.}}, \bibinfo {author} {\bibnamefont {Licata}}, \bibinfo
  {author} {\bibnamefont {I.}}, \bibinfo {author} {\bibnamefont {Moradpour}},\
  and\ \bibinfo {author} {\bibfnamefont {C.}~\bibnamefont {H.}, \bibfnamefont
  {Corda}},\ }\bibfield  {title} {\bibinfo {title} {On the general relativistic
  framework of the {S}agnac effect},\ }\href
  {https://doi.org/10.1140/epjc/s10052-019-6692-9} {\bibfield  {journal}
  {\bibinfo  {journal} {The European Physical Journal C}\ }\textbf {\bibinfo
  {volume} {79}},\ \bibinfo {pages} {187} (\bibinfo {year} {2019})}\BibitemShut
  {NoStop}%
\bibitem [{\citenamefont {Vavry\v{c}uk}\ and\ \citenamefont
  {K\v{r}\'{i}\v{z}ek}(2024)}]{pVav24}%
  \BibitemOpen
  \bibfield  {author} {\bibinfo {author} {\bibfnamefont {V.}~\bibnamefont
  {Vavry\v{c}uk}}\ and\ \bibinfo {author} {\bibfnamefont {M.}~\bibnamefont
  {K\v{r}\'{i}\v{z}ek}},\ }\bibfield  {title} {\bibinfo {title} {Symmetric twin
  paradox for free-falling frames: {A}rgument against the relativistic time
  dilation?},\ }\href
  {https://doi.org/https://doi.org/10.1016/j.physleta.2024.129886} {\bibfield
  {journal} {\bibinfo  {journal} {Physics Letters A}\ }\textbf {\bibinfo
  {volume} {525}},\ \bibinfo {pages} {129886} (\bibinfo {year}
  {2024})}\BibitemShut {NoStop}%
\bibitem [{\citenamefont {Lambare}(2025)}]{pLam25b}%
  \BibitemOpen
  \bibfield  {author} {\bibinfo {author} {\bibfnamefont {J.~P.}\ \bibnamefont
  {Lambare}},\ }\bibfield  {title} {\bibinfo {title} {Comment on ``{S}ymmetric
  twin paradox for free-falling frames: {A}rgument against the relativistic
  time dilation?''},\ }\href
  {https://doi.org/https://doi.org/10.1016/j.physleta.2025.130579} {\bibfield
  {journal} {\bibinfo  {journal} {Physics Letters A}\ }\textbf {\bibinfo
  {volume} {550}},\ \bibinfo {pages} {130579} (\bibinfo {year}
  {2025})}\BibitemShut {NoStop}%
\bibitem [{\citenamefont {Klauber}(1999)}]{pKla99}%
  \BibitemOpen
  \bibfield  {author} {\bibinfo {author} {\bibfnamefont {R.~D.}\ \bibnamefont
  {Klauber}},\ }\bibfield  {title} {\bibinfo {title} {{Comments regarding
  recent articles on relativistically rotating frames}},\ }\href
  {https://doi.org/10.1119/1.19213} {\bibfield  {journal} {\bibinfo  {journal}
  {American Journal of Physics}\ }\textbf {\bibinfo {volume} {67}},\ \bibinfo
  {pages} {158} (\bibinfo {year} {1999})}\BibitemShut {NoStop}%
\bibitem [{\citenamefont {Weber}(1999)}]{pWeb99}%
  \BibitemOpen
  \bibfield  {author} {\bibinfo {author} {\bibfnamefont {T.~A.}\ \bibnamefont
  {Weber}},\ }\bibfield  {title} {\bibinfo {title} {Response to ``{C}omments
  regarding recent articles on relativistically rotating frames'' [{A}m {J}.
  {P}hys. 67 (2), 158 (1999)]},\ }\href {https://doi.org/10.1119/1.19214}
  {\bibfield  {journal} {\bibinfo  {journal} {American Journal of Physics}\
  }\textbf {\bibinfo {volume} {67}},\ \bibinfo {pages} {159} (\bibinfo {year}
  {1999})}\BibitemShut {NoStop}%
\bibitem [{Note4()}]{Note4}%
  \BibitemOpen
  \bibinfo {note} {The fact that some find the \protect \textit {clock
  hypothesis} an independent assumption, \cite {bFoc66,pRos78,bTor96} while
  others a consequence of the theory that transforms it into a theorem \cite
  {pSfarti17,pFletcher13} is an example of the duality between considering the
  HL an independent assumption or a natural consequence of the
  theory.}\BibitemShut {Stop}%
\bibitem [{\citenamefont {M\o{}ller}(1952)}]{bMol52}%
  \BibitemOpen
  \bibfield  {author} {\bibinfo {author} {\bibfnamefont {C.}~\bibnamefont
  {M\o{}ller}},\ }\href@noop {} {\emph {\bibinfo {title} {The Theory of
  Relativity}}}\ (\bibinfo  {publisher} {Oxford,: Clarendon Press},\ \bibinfo
  {year} {1952})\BibitemShut {NoStop}%
\bibitem [{\citenamefont {Weber}(1997)}]{pWeb97}%
  \BibitemOpen
  \bibfield  {author} {\bibinfo {author} {\bibfnamefont {T.~A.}\ \bibnamefont
  {Weber}},\ }\bibfield  {title} {\bibinfo {title} {{Measurements on a rotating
  frame in relativity, and the {W}ilson and {W}ilson experiment}},\ }\href
  {https://doi.org/10.1119/1.18696} {\bibfield  {journal} {\bibinfo  {journal}
  {American Journal of Physics}\ }\textbf {\bibinfo {volume} {65}},\ \bibinfo
  {pages} {946} (\bibinfo {year} {1997})}\BibitemShut {NoStop}%
\bibitem [{\citenamefont {Fock}(1966)}]{bFoc66}%
  \BibitemOpen
  \bibfield  {author} {\bibinfo {author} {\bibfnamefont {V.}~\bibnamefont
  {Fock}},\ }\href@noop {} {\emph {\bibinfo {title} {The {T}heory of {S}pace,
  {T}ime \& {G}ravitation}}},\ \bibinfo {edition} {2nd}\ ed.\ (\bibinfo
  {publisher} {Pergamon Press Ltd.},\ \bibinfo {address} {Headington Hill Hall,
  Oxford OX3 0BW, England},\ \bibinfo {year} {1966})\BibitemShut {NoStop}%
\bibitem [{\citenamefont {Rosser}(1978)}]{pRos78}%
  \BibitemOpen
  \bibfield  {author} {\bibinfo {author} {\bibfnamefont {W.~G.~V.}\
  \bibnamefont {Rosser}},\ }\bibfield  {title} {\bibinfo {title} {The clock
  hypothesis and the lorentz transformations},\ }\href@noop {} {\bibfield
  {journal} {\bibinfo  {journal} {The British Journal for the Philosophy of
  Science}\ }\textbf {\bibinfo {volume} {29}},\ \bibinfo {pages} {349}
  (\bibinfo {year} {1978})}\BibitemShut {NoStop}%
\bibitem [{\citenamefont {Torretti}(1996)}]{bTor96}%
  \BibitemOpen
  \bibfield  {author} {\bibinfo {author} {\bibfnamefont {R.}~\bibnamefont
  {Torretti}},\ }\href {https://books.google.com.py/books?id=vpW_sBxwr88C}
  {\emph {\bibinfo {title} {Relativity and {G}eometry}}},\ Dover Books on
  Physics Series\ (\bibinfo  {publisher} {Dover Publications, Incorporated},\
  \bibinfo {address} {31 East 2nd Street, Mineola, N.Y. 11501},\ \bibinfo
  {year} {1996})\BibitemShut {NoStop}%
\bibitem [{\citenamefont {Sfarti}(2017)}]{pSfarti17}%
  \BibitemOpen
  \bibfield  {author} {\bibinfo {author} {\bibfnamefont {A.}~\bibnamefont
  {Sfarti}},\ }\bibfield  {title} {\bibinfo {title} {General {C}locks and the
  {C}lock {H}ypothesis},\ }\href
  {https://doi.org/https://dx.doi.org/10.22606/tp.2017.24005} {\bibfield
  {journal} {\bibinfo  {journal} {Theoretical Physics}\ }\textbf {\bibinfo
  {volume} {2}},\ \bibinfo {pages} {188} (\bibinfo {year} {2017})}\BibitemShut
  {NoStop}%
\bibitem [{\citenamefont {Fletcher}(2013)}]{pFletcher13}%
  \BibitemOpen
  \bibfield  {author} {\bibinfo {author} {\bibfnamefont {S.~C.}\ \bibnamefont
  {Fletcher}},\ }\bibfield  {title} {\bibinfo {title} {Light {C}locks and the
  {C}lock {H}ypothesis},\ }\href
  {https://doi.org/https://doi.org/10.1007/s10701-013-9751-3} {\bibfield
  {journal} {\bibinfo  {journal} {Foundations of Physics}\ }\textbf {\bibinfo
  {volume} {43}},\ \bibinfo {pages} {1369} (\bibinfo {year}
  {2013})}\BibitemShut {NoStop}%
\end{thebibliography}%
\end{document}